\documentclass[fleqn]{2017SCGE}
\begin{document}

\ensubject{subject}

%%%%%%%%%%%%%%%%%%%%%%%%%%%%%%%%%%%%%%%%%%%%%%%%%%%%%%%
%%% Authors do not modify the information below
%%% ????????????????
%%% ??????????, ????????????{}, ???????????????????
%Letter to the Editor??Article%??????
\ArticleType{Article}%??Article
\SpecialTopic{SPECIAL TOPIC: FAST special issue}%???????
\Year{2019}
\Month{May}
\Vol{62}
\No{5}
\DOI{xxx}
\ArtNo{959504}
\ReceiveDate{December 6, 2018}
\AcceptDate{January 19, 2019}
%\OnlineDate{January 1, 2016}
%%%%%%%%%%%%%%%%%%%%%%%%%%%%%%%%%%%%%%%%%%%%%%%%%%%%%%%

%%% title: ????
%%%   \title{title}{title for citation}
\title{FAST ultra-wideband observation of abnormal emission-shift events of PSR B0919$+$06}{FAST ultra-wideband observation of abnormal emission-shift events of PSR B0919$+$06}

%%% Corresponding author: ???????
%%%   \author[number]{Full name}{{email@xxx.com}}
%%% General author: ???????
%%%   \author[number]{Full name}{}
\author[1,2]{Ye-Zhao Yu}{{yzyu@nao.cas.cn}}%
\author[1]{Bo Peng}{pb@nao.cas.cn}
\author[3,1]{Kuo Liu}{}
\author[1]{\\Chengmin Zhang}{}
\author[1,2]{Lin Wang}{}
\author[1]{Fei Fei Kou}{}
\author[1]{Jiguang Lu}{}
\author[1]{Meng Yu}{}
\author[1]{\\FAST Collaboration}{}

%%% Author information for page head. ?¨¹?§Ö????????
%%% ??????????????, ??????????author???
\AuthorMark{Y.-Z.~Yu, B.~Peng, K.~Liu, et al}%\authorcr????????

%%% Authors for citation. ????????§Ö????????
%%% ??????????????, ??????????author???
\AuthorCitation{Y.-Z.~Yu, B.~Peng, K.~Liu, et al}

%%% Address. ???
%%%   \address[number]{Address, City {\rm Postcode}, Country}
\address[1]{CAS Key Laboratory of FAST, National Astronomical Observatories, Chinese Academy of Sciences, Beijing {\rm 100101}, China}
\address[2]{College of Astronomy and Space Sciences, University of Chinese Academy of Sciences, Beijing {\rm 100049}, China}
\address[3]{Max-Planck-Institut f$\ddot{u}$r Radioastronomie, Auf dem H$\ddot{u}$gel 69, {\rm 53121} Bonn, Germany}

%\contributions{}%????????

%%% Abstract. ??
\abstract{PSR~B0919$+$06 is known for its abnormal emission phenomenon, where the pulse emission window occasionally shifts progressively in longitude and returns afterwards. The physical mechanism behind this phenomenon is still under investigation. In this paper, we present our ultra-wideband observation of this pulsar using the Five-hundred-meter Aperture Spherical radio Telescope (FAST), with simultaneous measurements in the frequency ranges 280--780\,MHz and 1250--1550\,MHz.
We have identified three abnormal events, each of which becomes less apparent as the frequency decreases. At 1400\,MHz, the averaged profile slightly shifted after the first and third abnormal events, implying a relationship between abnormal event and profile variation. We also found a linear trend in the left-edge position of the averaged profiles between the first and third event as well as after the third event, suggesting the existence of a slow-drifting mode between the two major events. The second event has a comparatively small shift in phase and is thus categorized as a ``small flare state''. During the third event, a sequence of approximately nine pulses was seen to significantly weaken in all frequency bands, likely associated with the pseudo-nulling observed at 150\,MHz. A three-component de-composition analysis of the normal averaged profiles shows that the trailing component is dominant at our observing frequencies, while the centre component has a comparatively steeper spectrum.
We found the overall flux density in an abnormal event to slightly differ from that in an ordinary state, and the difference shows a frequency dependence.
A comparison of the normal, abnormal and dimmed averaged profile indicates that the leading component is likely to be stable in all states.}%ÕªÒª

%%% Keywords. ?????
%stars: neutron --- pulsars: general --- pulsars: individual: PSR B0919$+$06


\maketitle

%\tableofcontents%?????

%%%%%%%%%%%%%%%%%%%%%%%%%%%%%%%%%%%%%%%%%%%%%%%%%%%%%%%
%%% The main text. ???????
%???????????????????\cref{fig1}
%\twocolumn\onecolumn
%%%%%%%%%%%%%%%%%%%%%%%%%%%%%%%%%%%%%%%%%%%%%%%%%%%%%%%
\begin{multicols}{2}
\section{Introduction}
\label{sect:intro}
Pulsars are rapidly rotating magnetized neutron stars, the radio pulsation signals from which are observed when the radiation beam passes across the observer's line of sight. The classical phenomena of variations of individual pulses or averaged profiles, e.g. subpulse drifting, where the
\Authorfootnote

\noindent
subpulses periodically drift in longitude in the pulsar emission window~\cite{1970Natur.227..692B, 2006A&A...445..243W, 2007A&A...469..607W}; nulling, where the energy of an individual pulse incidentally drops to zero or near zero~\cite{1970Natur.228...42B, 1992ApJ...394..574B, 2007MNRAS.377.1383W}; and mode changing where the integrated profiles switch between two or more profile forms~\cite{1970Natur.228.1297B, 1982ApJ...258..776B, 2007MNRAS.377.1383W}, have been detected in many pulsars. An additional variation phenomenon, in which the individual pulses first shift in longitude toward early phase and then turn backward progressively, was first identified by ref.~\cite{2006MNRAS.370..673R} in PSR~B0919$+$06 and B1859$+$07. Such an abnormal emission event is difficult to categorize within the classical categories, and its physical mechanism is still unknown.

PSR~B0919$+$06 is a bright pulsar with a rotation period of 0.4306\,s~\cite{2013ApJ...775....2S} and a dispersion measure of 27.2986\,pc\,$\rm cm^{-3}$~\cite{2015ApJ...808..156S}.
Its averaged profile is significantly asymmetric in the frequency range from 61 to 1400\,MHz with a long weak leading component and a bright trailing component.
An additional central component can be detected clearly within the frequency range 300--500\,MHz (according to the profiles in ref.~\cite{2006MNRAS.370..673R, 2008MNRAS.388..261J, 2010AJ....139..168H}). The abnormal emission events of PSR~B0919$+$06 were first discovered by ref.~\cite{2006MNRAS.370..673R} using the Arecibo 305-m telescope at 327 and 1400\,MHz. In such an abnormal event, the trailing subpulses occasionally shift progressively in longitude toward early phase and return later. Unlike the observations at 1400\,MHz, the abnormal event is hardly discernible in the total intensity at 327\,MHz, but it can still be identified by linear/circular polarization or from the rotation of the polarization angle. The abnormal events of PSR~B0919$+$06 were also identified in observations by ref.~\cite{2015MNRAS.446.1380P} using the Lovell 76-m telescope at 1400\,MHz. In addition to events that were termed ``main flare states'', they also found ``small flare states'', wherein pulses shift only by about $3^\circ$.
Ref.~\cite{2016MNRAS.456.3413H}~observed PSR~B0919$+$06 for about 30\,h using the Jimusi 66-m telescope at 2250\,MHz.
They detected 92 abnormal events and classified them into four morphological types, which they called $\Pi$, $\mathrm{M}$, $\Lambda$ and $\lambda$.
Ref.~\cite{2016MNRAS.461.3740W} searched for periodicity in the abnormal events using datasets from ref.~\cite{2006MNRAS.370..673R}, ref.~\cite{2016MNRAS.456.3413H}, and from additional extensive Arecibo observations.
In the Arecibo observations, among 26 abnormal events, they found an approximate period of about 700 pulses, or 300\,s, but no indication of periodicity was found in the dataset of ref.~\cite{2016MNRAS.456.3413H} with the 96 abnormal events. Ref.~\cite{2018MNRAS.477L..25S} observed PSR~B0919$+$06 by simultaneously using the Effelsberg 100-m telescope at 1350\,MHz and the Bornim station of the LOw Frequency ARray (LOFAR) at 150\,MHz.
They detected seven abnormal events in their 1350\,MHz observations, but they observed no emission shift at 150\,MHz.
Instead, on most occasions the intensity of 10-s sub-integrations at 150\,MHz decreased as `pseudo-nulling' when an abnormal event occurred at 1350\,MHz.
As no significant intensity variation of the sub-integrations was observed at 1350\,MHz, they considered that such  pseudo-nulling only occurs at a low frequency. There have been no previous observations in the frequency range 327--1350\,MHz for the abnormal events of PSR~B0919$+$06.

In this article, we present our ultra-wideband observation of PSR~B0919$+$06, covering the frequency ranges 280--780 and 1250--1550\,MHz, respectively.
The observation is described in \cref{sect:obs}. The results are shown in \cref{sect:result}, and discussions are placed in \cref{sect:discussion}.
Finally, a summary is given in \cref{sect:summary}.

\section{Observation}
\label{sect:obs}
The Five-hundred-meter Aperture Spherical radio Telescope (FAST)~\cite{2001Ap&SS.278..219P, 2006ScChG..49..129N} is located at Pingtang, Guizhou Province, China. It was built and is operated by the National Astronomical Observatories, Chinese Academy of Sciences (NAOC).
Its construction was completed in September, 2016, and the telescope is currently in the commissioning phase~\cite{2019ScChG..J}. FAST has total and effective apertures of 500\,m and 300\,m, respectively. An ultra-wideband receiver covering 270 to 1620\,MHz was used between September 2016 and May 2018.
The observation presented in this article was obtained on 8th October 2017 (MJD~58034), with a duration of approximately 30\,min. The signal collected by the receiver was filtered into two individual bands with frequency range of 270 to 800\,MHz and 1200 to 1620\,MHz, respectively. Next, we used a digital backend based on Reconfigurable Open Architecture Computing Hardware generation 2 (ROACH 2)\footnote{https://casper.berkeley.edu/} to sample each band at the Nyquist rate, yield intensity detections for dual polarisations, and pack the data in PSRFITS~\cite{2004PASA...21..302H} search-mode format with 0.2\,ms samples and 0.25\,MHz frequency channels.

We then processed the data with the \textsc{DSPSR} software~\cite{2011PASA...28....1V} to generate single-pulse archives, each with 2048 phase bins. In total, we obtained 3916 single pulses from PSR~B0919$+$06. Channels containing radio frequency interferences were identified by eye and removed manually.
Only the data in the frequency ranges 280--780 and 1250--1550\,MHz were used after this step.
For each frequency channel and polarisation, we also subtracted the baseline and re-scaled the data by the amplitude of the noise diode from observation of the ``clear sky'', which mitigated the amplitude differentials between the two polarisations.
%For each frequency channel and polarisation, we also re-scaled the data according to the observation for `clear sky' with the noise diode switch between on and off. In addition, the low band data were split into five 100-MHz sub-bands centred at 330, 430, 530, 630 and 730\,MHz.
Finally, we averaged the data in the frequency domain to improve the signal-to-noise ratio for the detection of single pulses.

\section{Result}
\label{sect:result}
\begin{figure*}
	\centering
	\includegraphics[width=17.5cm, angle=0]{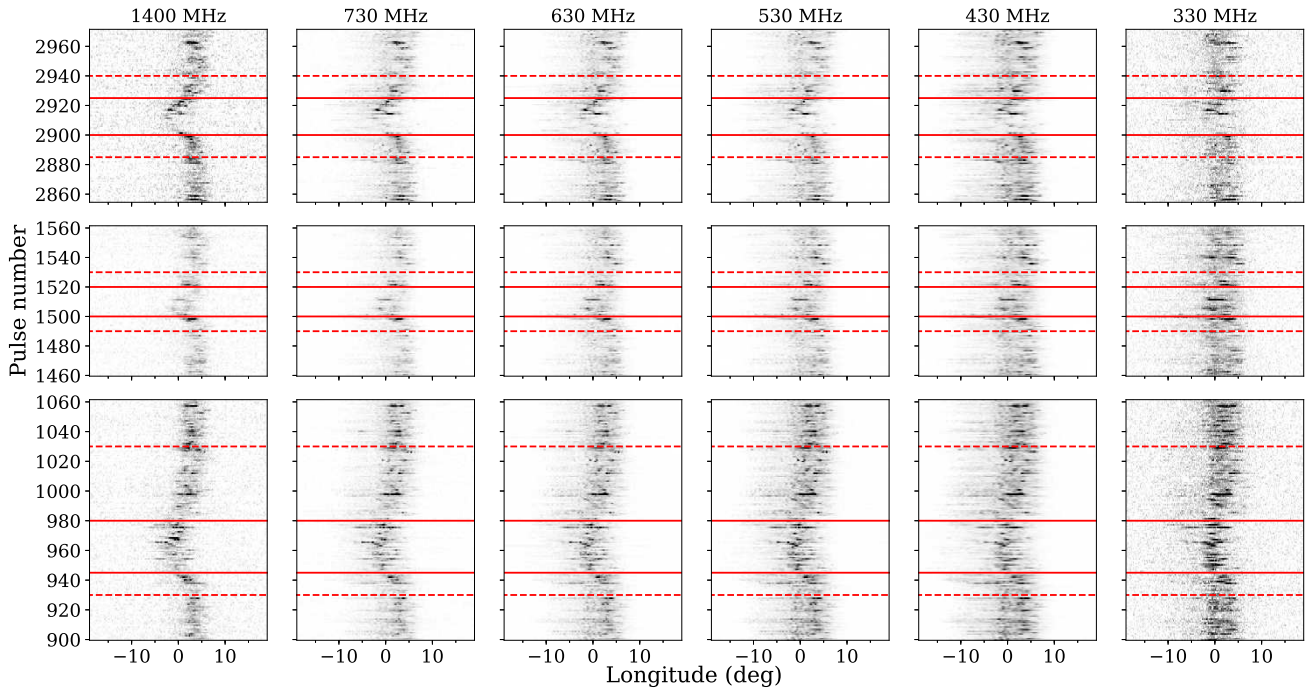}
	\caption{The three abnormal events identified in our observation. Panels in the same row show the same abnormal event in different frequency bands. Panels in the same column belong to the same frequency band, for which the centre frequency is marked at the top. Pulses between red dashed lines belong to the abnormal event, while the range constrained by the red solid lines are the regions with the ``most-shifted'' pulses.}
	\label{fig:events}
\end{figure*}

\begin{figure}[H]
	\centering
	\includegraphics[width=7.6cm, angle=0]{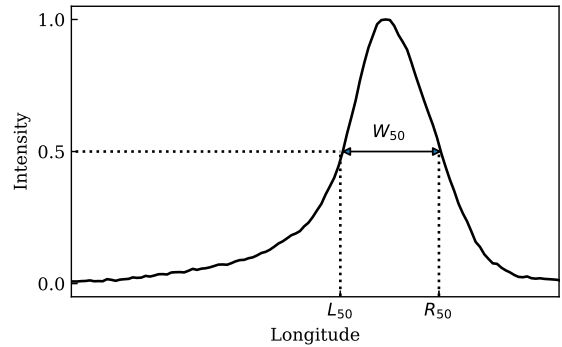}
	\caption{An example showing the definitions of $L_{\rm 50}$, $R_{\rm 50}$, and $W_{\rm 50}$. The black solid line shows a normalised integrated profile of PSR B0919$+$06.}
	\label{fig:example}
\end{figure}

\begin{figure*}
	\centering
	\includegraphics[width=15.5cm, angle=0]{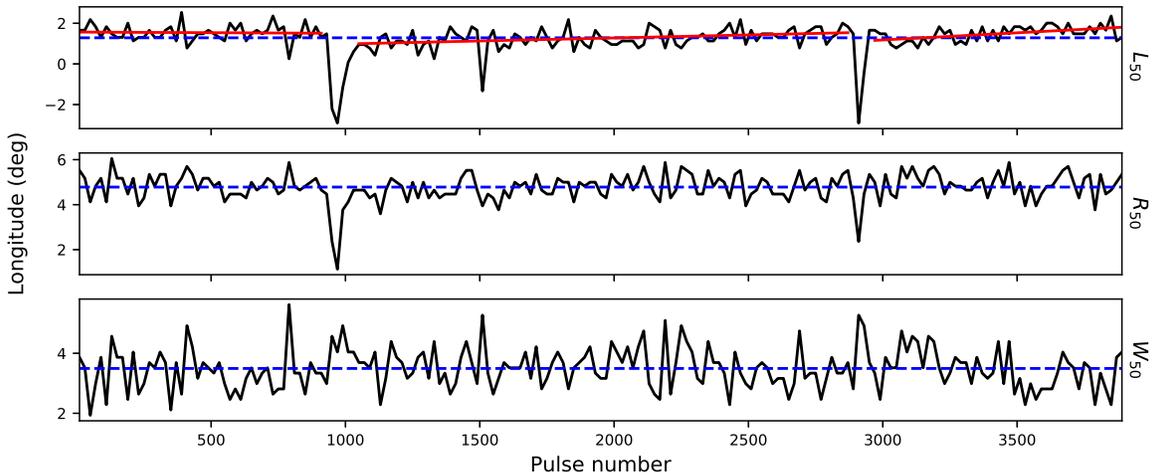}
	\caption{Measurements of $L_{\rm 50}$, $R_{\rm 50}$, and $W_{\rm 50}$ from twenty-period sub-integrations are shown in the upper, middle and lower panels, respectively. The blue dashed lines represent the overall mean values. The red lines in the upper panel are linear fits to the $L_{\rm 50}$ values in three intervals: before the first event, between the first and third event, and after the third event. The obtained slopes are in order, $(-0.62\pm 2.28)\times10^{-4}$, $(2.96\pm 0.755)\times 10^{-4}$, and $(7.02\pm 1.74)\times 10^{-4}\,{\rm deg}/P_{0}$, where $P_{\rm 0}$ is the pulse period.}
	\label{fig:W_50}
\end{figure*}

We identified three abnormal events in our observation. \cref{fig:events} shows the individual pulse sequences of these three events in the different frequency bands. To identify the time intervals containing the abnormal events, we first averaged individual pulses from the 1400\,MHz observation to produce twenty-period (about 8.6\,s) sub-integrations. For each sub-integration, we then measured the phase of the leftmost (earliest) and rightmost (latest) bin that have intensities larger than 50\% of the peak ($L_{\rm 50}$ and $R_{\rm 50}$, respectively); the width at 50\% of the peak is $W_{\rm 50}=R_{\rm 50}-L_{\rm 50}$.
An example illustrating the measurement of $L_{\rm 50}$, $R_{\rm 50}$, and $W_{\rm 50}$ is shown in \cref{fig:example}. The values of $L_{\rm 50}$, $R_{\rm 50}$, and $W_{\rm 50}$ are plotted in \cref{fig:W_50}. Based on the significant decrease of the $L_{\rm 50}$ values at about pulse numbers 1000, 1500, and 2900, we identified the regions of the three abnormal events where all $L_{\rm 50}$ values are below average. Within the region of each event, we have also highlighted the range where $L_{\rm 50}$ is below 50\% of the minimum $L_{\rm 50}$ in that event. Pulses from these ranges are later referred to as ``most-shifted'' pulses. The first and third events exhibit a shift in longitude up to $5^\circ$, while the second event has a maximum shift of only $3^\circ$ and may fall into the category of ``small flare state'', first discussed by ref.~\cite{2015MNRAS.446.1380P}. The results of the $L_{\rm 50}$, $R_{\rm 50}$, and $W_{\rm 50}$ measurements are summarized in \cref{tab:pulse}.

\begin{table}[H]
	%\centering
	\footnotesize
	\caption{Pulse numbers of the abnormal events in our observation.}
	\label{tab:pulse}
	\begin{tabular}{ccc}\hline
		Event & Entire event & Most-shifted \\
		no.   & pulse no. & pulse no. \\\hline
		1   &    930--1030    &    945--980 \\
		2   &    1490--1530   &    1500--1520 \\
		3   &    2885--2940   &    2900--2925 \\\hline
	\end{tabular}
\end{table}

A close observation of the pulse sequence at 1400\,MHz in \cref{fig:events} indicates that the left edges of the individual pulses after the first abnormal event are still slightly offset from the previous level before the event.
The same phenomenon was noted before and after the third event as well. This observation is supported by the measured $L_{\rm 50}$ values for the sub-integrations shown in \cref{fig:W_50}, where the percentage of measurements above average drops from 85\% to 35\% and from 80\% to 35\%, after the first and the third event, respectively. The value of $L_{\rm 50}$ exhibits an increase in time, both from the end of the first event until the beginning the third event and from the end of the third event onwards. As shown in \cref{fig:W_50}, linear fits to the $L_{\rm 50}$ values in these two intervals verify the significance of both trends.
This suggests the possible existence of a slow-drifting mode between the two major events, where the intensity shift is much more rapid. Further observations with more detections of abnormal events will be required to confirm this finding.

In all three cases, the abnormal event becomes less apparent as the observing frequency decreases. Such a frequency dependence of the abnormal events is in line with the expectation of ref.~\cite{2006MNRAS.370..673R}, who found that the abnormal event is difficult to identify in the total-intensity pulse sequence at low frequency. A similar phenomenon has also been witnessed by ref.~\cite{2018MNRAS.477L..25S}. However, it is noteworthy that during the third abnormal event, there are approximately nine individual pulses (pulse numbers 2902 to 2910) that appear to be dimmer than the others (see the upper panels in \cref{fig:events}). Herein, shifted sub-pulses are barely visible in any frequency band, although a weak but broad component can been seen in the 430 to 730\,MHz bands that does not exhibit an apparent shift in time.

\begin{figure}[H]
	\centering
	\includegraphics[width=7cm, angle=0]{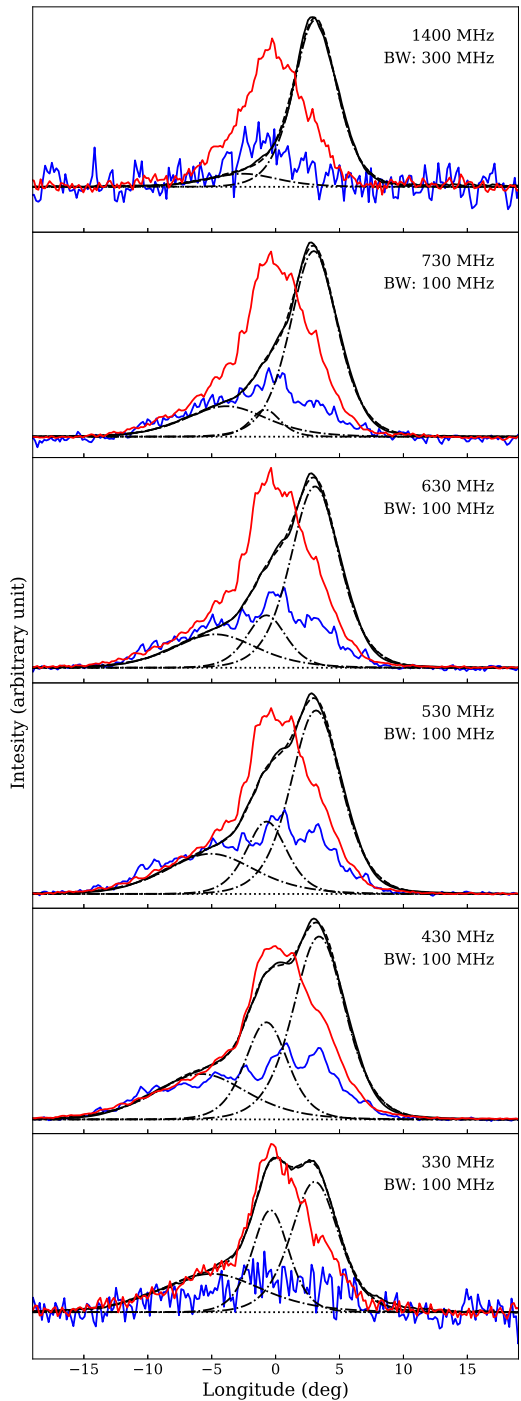}
	\caption{Averaged profiles for normal, obvious pulse-phase shifted and dimmed pulses. The different panels display profiles in different frequency bands with the centre frequency and bandwidth (BW) marked at the top right. Black solid lines are the profiles averaged over individual pulses excluding those between the red dashed lines in \cref{fig:events}. The red solid lines are averaged profiles of the most-shifted pulses, which are in the range between the red solid lines in \cref{fig:events}. Averaged profiles of the dimmed pulses in the third abnormal pulses are plotted using blue solid lines. The zero intensity in each panel is indicated by a black dotted line. The de-composition results of the normal profile at each frequency band are shown in black dash-dotted lines for the profile components and as a black dashed line for the whole profile.}
	\label{fig:profiles}
\end{figure}

The averaged profiles of normal, abnormal (most-shifted pulses only) and dimmed pulses in different frequency bands are shown in \cref{fig:profiles}. At 1400\,MHz and 330\,MHz, the normal and abnormal averaged profiles are consistent with those obtained by ref.~\cite{2006MNRAS.370..673R}. For the normal averaged profiles, the intensity of the centre component (peaked around $0^\circ$ longitude) seems to drop significantly relative to the trailing component and is barely detectable at 1400\,MHz, suggesting that the centre component has a significantly steeper spectrum than other components. The overall trend also indicates that the main peak and the plateau at the trailing side of the averaged profile observed in ref.~\cite{2018MNRAS.477L..25S} at 150\,MHz correspond to the centre and the trailing component seen herein, respectively.
To further investigate the variation in frequency, we performed a three-component de-composition of the normal averaged profiles in all frequency bands. We found that the trailing component is dominant across the entire frequency range. The centre component becomes weaker than the leading component at 730\,MHz, and it is insignificant in the fit at 1400\,MHz.

The abnormal averaged profiles approximately peak at the longitude of the centre component, and they exhibit less apparent variation across the frequency bands. Compared with the flux densities of the normal averaged profile at the same frequency, their flux densities in general are at the same level, but they are slightly stronger at 1400\,MHz and become relatively weaker as the frequency decreases. This indicates a small flattening in the overall spectrum during the abnormal event. At 330 and 430\,MHz, the overlap of the normal and abnormal profiles at the leading edge, i.e. at negative longitudes, indicates that both the leading and the centre component are unlikely to shift during the abnormal event. The amplitude of the centre component, as indicated from the de-composition, is significantly lower than the peak of the abnormal averaged profile. This means that additional emission occurs at the longitude of the centre component during the abnormal event. All three profiles overlaps around the region of the leading component from 430\,MHz to 730\,MHz, suggesting that the leading component mostly remains unchanged during the abnormal event. In addition, the profiles at 330 and 430\,MHz shows that both the centre and trailing components become weaker during the dimming process.

\section{Discussion}
\label{sect:discussion}
The profile parameters ($L_{\rm 50}$, $R_{\rm 50}$ and $W_{\rm 50}$) of the sub-integrations obtained at 1400\,MHz evidence the profile variation before and after the abnormal event.
We also noticed that the left edges of the pulses after the first abnormal event, as shown in Figure~10 of ref.~\cite{2015MNRAS.446.1380P}, also arrive at an earlier phase than those before the event. Nevertheless, additional observations of PSR~B0919$+$06 are required to gather more information about the relationship between the abnormal event and profile variations.

Compared with the pseudo-nulling phenomenon identified in ref.~\cite{2018MNRAS.477L..25S}, the dimming of pulses seen in our observation has a shorter duration ($\lesssim10$\,s) and occurs over a wide frequency range. On the other hand, the dimmed averaged profile at 330\,MHz is similar to that at 150\,MHz in their work. Given the fact that the observations in ref.~\cite{2018MNRAS.477L..25S} have a limited time resolution (10\,s), it is still possible that our dimmed pulses are in fact the same phenomenon. This, however, needs to be investigated with further observations.

A number of possible origins of the abnormal events of PSR~B0919$+$06 have been proposed. Ref.~\cite{2006MNRAS.370..673R}~suggested that the abnormal event may be an extreme example of variable absorption within the pulsar magnetosphere or that it may be due to intrinsic changes of the emission region in the pulsar beam.
Ref.~\cite{2018MNRAS.477L..25S} claimed support for the absorption model, although they noted that the absorption needs to have frequency-dependent efficiency to explain the pseudo-nulling phenomenon that only occurs at 150\,MHz. However, since we have found that the pseudo-nulling phenomenon occurs over a wide frequency range, there is no longer a need for proving such a dependence.
The abnormal event could also be caused by intrinsic relocations of the emission spots within the pulsar beam.
Ref.~\cite{2017MNRAS.469.2049Y}~applied their differentially co-rotating magnetosphere model to the abnormal events of PSR~B0919$+$06, and were able to reproduce the emission shift at 1400\,MHz.
If the emission spots move both along and away from the line of sight, we might also observe the wide-band pseudo-nulling inside an abnormal event.
Due to the discovery of a possible quasi-periodicity of the abnormal event in the Arecibo observations, ref.~\cite{2016MNRAS.461.3740W} and ref.~\cite{2018ApJ...855...35G} suggested that the abnormal event may be caused by a 'companion' moving around the pulsar in a highly eccentric orbit.
However, as no apparent emission shift of the leading or centre component has been observed in our data
and because the pulse profile exhibits consistent polarization angle behavior between the normal and abnormal state, as shown and discussed in ref.~\cite{2006MNRAS.370..673R} and ref.~\cite{2016MNRAS.456.3413H},
the abnormal event is unlikely to be associated with a binary system.

\section{Summary}
\label{sect:summary}
We have presented ultra-wideband observation of PSR~B0919$+$06 using FAST, with simultaneous measurements in the frequency ranges 280--780 and 1250--1550\,MHz. We identified three abnormal events in our observation which contains 3916 individual pulses. The averaged profile at 1400\,MHz shifts slightly after the first and the third abnormal event, indicating a relationship between the abnormal event and profile changing. We also found a linear trend in the position of the left edge of the averaged profile both between the first and third event and after the third, indicating the possible existence of a slow-drifting mode between these two major events. The second event has a smaller pulse-phase shift and is thus likely to be categorized as a ``small flare state''. In the third event, a sequence of approximately nine pulses was seen to become significantly dimmer in all frequency bands, which later has been associated with the pseudo-nulling observed at 150\,MHz.

A three-component de-composition of the normal averaged profile shows that the trailing component is dominant across our observing frequencies, while the centre component has a comparatively steeper spectrum and fades away at 1400\,MHz. During the abnormal event, the overall flux density generally remains the same relative to that of the normal state but shows slight variations at different frequencies. A comparison of the normal, abnormal, and dimmed averaged profiles indicates that the leading component is likely to remain mostly unchanged during the abnormal event.

%%%%%%%%%%%%%%%%%%%%%%%%%%%%%%%%%%%%%%%%%%%%%%%%%%%%%%%
%%% Acknowledgements. ??§Ý
%%%%%%%%%%%%%%%%%%%%%%%%%%%%%%%%%%%%%%%%%%%%%%%%%%%%%%%
\Acknowledgements{
This work made use of the data from the FAST telescope (Five-hundred-meter Aperture Spherical radio Telescope). FAST is a Chinese national mega-science facility, built and operated by the National Astronomical Observatories, Chinese Academy of Sciences.
This work is supported by the National Natural Science Foundation of China (No. 11673031, No. 11703048 and U1731238), and the Open Project Program of the Key Laboratory of FAST, NAOC.
KL acknowledges the financial support by the European Research Council for the ERC Synergy
Grant BlackHoleCam under contract no. 610058, the FAST FELLOWSHIP from Special Funding for Advanced Users, budgeted and administrated by Center for Astronomical Mega-Science, Chinese Academy of Sciences (CAMS), and the MPG-CAS Joint Project ``Low-Frequency Gravitational Wave Astronomy and Gravitational Physics in Space''.
}

%%%%%%%%%%%%%%%%%%%%%%%%%%%%%%%%%%%%%%%%%%%%%%%%%%%%%%%
%%% Conflict of interest. ????????????
%%%%%%%%%%%%%%%%%%%%%%%%%%%%%%%%%%%%%%%%%%%%%%%%%%%%%%%
%\InterestConflict{The authors declare that they have no conflict of interest.}

%%%%%%%%%%%%%%%%%%%%%%%%%%%%%%%%%%%%%%%%%%%%%%%%%%%%%%%
%%% Supplements. ????????, ????
%%%%%%%%%%%%%%%%%%%%%%%%%%%%%%%%%%%%%%%%%%%%%%%%%%%%%%%
%\Supplements{}

%%%%%%%%%%%%%%%%%%%%%%%%%%%%%%%%%%%%%%%%%%%%%%%%%%%%%%%
%%% Reference section. ?¦Ï?????
%%% citation in the content using "some words~\cite{1,2}".
%%% ~ is needed to make the reference number is on the same line with the word before it.
%%%%%%%%%%%%%%%%%%%%%%%%%%%%%%%%%%%%%%%%%%%%%%%%%%%%%%%

\end{multicols}
\end{document}